\documentclass[citeauthoryear]{llncs}

\usepackage{graphicx}
\usepackage[cmex10]{amsmath}
\usepackage{amsfonts}
\usepackage{amssymb}
\usepackage{mathtools}
\usepackage{bbm}
\usepackage{xspace}
\usepackage{xparse}
\usepackage{subfigure}
\usepackage{booktabs}
\usepackage{adjustbox}
\usepackage{pbox}
\usepackage{cleveref}

\usepackage[backend=biber,bibencoding=utf8]{biblatex}

\bibliography{ms.bib}

\crefname{figure}{Figure}{Figures}%
\crefname{table}{Table}{Tables}%
\crefname{section}{Section}{Sections}%

\newcommand{\termdef}[1]{\mbox{\emph{#1}}}

\newcommand{\vertexnotation}[1]{$v_{#1}$\xspace}
\newcommand{\vertex}[1]{\vertexnotation{#1}}
\newcommand{\vertices}[2]{\vertexnotation{#1} and~\vertexnotation{#2}}
\newcommand{\betweenvertices}[2]{\vertices{#1}{#2}}

\newcommand{\edgenotation}[2]{\mbox{$(#1, #2)$}\xspace}
\newcommand{\edge}[2]{edge~\edgenotation{#1}{#2}}

\newcommand{\timenotation}[1]{\mbox{$#1$}\xspace}
\renewcommand{\time}[1]{time~\timenotation{#1}}

\newcommand{\class}[1]{class~\mbox{$#1$}\xspace}

\newcommand{\systemnotation}[1]{\mbox{$#1$}\xspace}
\newcommand{\system}[1]{system~\systemnotation{#1}}

\newcommand{\particle}[1]{particle~$p$\xspace}

\newcommand{\Classes}{\ensuremath{C}}
\newcommand{\PartClasses}{\ensuremath{K}}
\newcommand{\Comm}[1]{\ensuremath{y_{#1}}}

\newcommand{\DataPoints}{\mathcal{X}}

\newcommand{\suchthat}{|}
\newcommand{\card}[1]{\left\lvert#1\right\rvert}

\newcommand{\VertexSet}{\mathcal{V}}
\newcommand{\EdgeSet}{\mathcal{E}}

\newcommand{\WeightSet}{\mathcal{W}}

\newcommand{\ModelName}{Edge Domination System\xspace}

\newcommand{\WDynamicsName}{walking\xspace}
\NewDocumentCommand\MaxSet{m+g}{%
    \IfNoValueTF{#2}%
        {\max\!\left\{#1\right\}}%
        {\max\!\left\{#1,\;#2\right\}}%
}
\NewDocumentCommand\MinSet{m+g}{%
    \IfNoValueTF{#2}%
        {\min\!\left\{#1\right\}}%
        {\min\!\left\{#1,\;#2\right\}}%
}

\newcommand{\submission}[1][c]{\sigma^{#1}_{ij}}
\newcommand{\RelativeCurrentSubmission}[1][c]{\tilde{\sigma}^{#1}_{ij}}
\newcommand{\CurrentDomination}[2][ij]{\tilde{n}^{#2}_{#1}}

\newcommand{\FinalTrans}{P^c}
\newcommand{\finaltrans}[1][ij]{p^c_{#1}}

\newcommand{\SurvivalParam}{\lambda}

\newcommand{\Sourceness}{\rho^c_i}
\newcommand{\SourceSet}[1][c]{\mathcal{G}^c}

\newcommand{\onewparticles}[1][i]{\tilde{g}^c_{#1}}

\newcommand{\onpart}[1][ij]{\tilde{n}^c_{#1}}
\newcommand{\onpartclass}[2][ij]{\tilde{n}^{#2}_{#1}}

\newcommand{\ovpart}[1][c]{\vec{\tilde{n}}^c}
\newcommand{\npart}[1][ij]{n^c_{#1}}
\newcommand{\xnpart}[2][ij]{n^{#2}_{#1}}

\newcommand{\vpart}[1][c]{\vec{n}^{#1}}
\newcommand{\Npart}[1][c]{N^{#1}}

\DeclareMathOperator*{\argmax}{arg\,max}
\DeclareMathOperator*{\argmin}{arg\,min}
\newcommand{\diag}{\operatorname{diag}}

\newcommand{\defeq}{\coloneqq}

\newcommand{\BigO}{\mathcal{O}}
\newcommand{\Complexity}[1]{\BigO\!\left(#1\right)}

\newcommand{\kNN}{{\itshape k}-NN\xspace}

\title{ Data clustering with edge domination in complex networks}
\author{Paulo Roberto Urio\inst{1} \and Zhao Liang\inst{2}}

\institute{
        Institute of Mathematical and Computer Sciences \\
        University of S\~{a}o Paulo \\
        S\~{a}o Carlos, SP, Brazil \\
        \email{urio@usp.br} \and
        Ribeirão Preto School of Philosophy, Science and Literature \\
        University of S\~{a}o Paulo \\
        Ribeir\~{a}o preto, SP, Brazil
}

\begin{document}
  \maketitle

\begin{abstract}
  This paper presents a model for a dynamical system
  where particles dominate edges in a complex network.  The proposed dynamical
  system is then extended to an application on the problem of community
  detection and data clustering.  In the case of the
  data clustering problem, 6 different techniques were simulated on
  10 different datasets in order to compare with the proposed technique.
  The results show that the proposed algorithm performs well when
  prior knowledge of the number of clusters is known to the algorithm.
\end{abstract}

\newcommand{\apriori}{\emph{a priori}\xspace}
\newcommand{\kmeans}{\emph{K}-means\xspace}
\newcommand{\cmeans}{\emph{c}-means\xspace}
\newcommand{\hdbscan}{HDBSCAN*\xspace}
\newcommand{\hdbscandbcv}{\hdbscan + DBCV\xspace}

\section{Introduction}

Consider a dataset that is represented by a weighted, undirected graph where a
vertex represents a data point and an edge a relationship
of similarity.  Particularly, if a dynamical system can take place in this
graph representation then this graph can be studied with tools of
the complex network theory~\cite{Barrat2008}.

In machine learning, methods that are based on networks have increasingly being
studied.  The representation of a dataset as a network allows the method to
work not only with a similarity score among pairs of nodes, but it also
provides a topological information.  Within this context, the goal of this
paper is to present a dynamical system of
particle competition system on edges of complex networks for the community
detection and the data clustering problem.

Regarding previous work with semi-supervised learning, in this paper the presented
model modifies dynamics of particle generation and introduces the information
of weights in edges.  In \cref{sec:model} the model is introduced with an
overview and a mathematical description.  In \cref{sec:simulations} simulations
on a real network, artificial datasets, and real-world datasets are presented.
Finally in \cref{sec:conclusions}, some final considerations regarding the
studied model are discussed.

\section{\ModelName} \label{sec:model}

In this section, I give an introduction to the proposed technique,
namely \ModelName. First an overview and then its mathematical modeling.

\subsection{Overview}

Consider a complex network expressed by a simple, undirected, weighted
graph $G = (\VertexSet, \EdgeSet)$, where $\VertexSet$ is the set of vertices
and $\EdgeSet \subseteq \VertexSet \times \VertexSet$ is the set of edges.
If two vertices are considered similar, then they are connected by an edge.
%
%
Denote \edgenotation{i}{j} to be the edge between vertices \betweenvertices{i}{j},
which has a weight $\WeightSet(i, j) \in \mathbb{R}$.  The weights represent the
similarity between two vertices where a larger value indicates higher
similarity.
The graph that expresses this network is represented by the weighted adjacency
matrix \mbox{$W = (w_{ij})$} where $w_{ij} = w_{ji} = \WeightSet(i, j)$ if \vertex{i} is
connected to \vertex{j} and zero otherwise.
The process' result consists of $\PartClasses$ groups of vertices that are not
necessarily connected.
%

In this model, particles are the objects that flow within the network.
Every particle belongs to a class, defined at time of their creation.
After being released in any node of the network, a particle randomly walks the
network. The probability among adjacent vertices to be chosen as the next vertex
follows the distribution of the weights of the possible edges.  Consider a
particle that is in~\vertex{i}.  This particle decides to
move to \vertex{j} with probability
\[
  w_{ij} \cdot \left( \sum_{k=1}^{|\VertexSet|} w_{ik} \right)^{\!-1} \mbox{.}
\]

The particle's decision, however, does not imply it will succeed at
moving to the neighboring vertex.  If the edge that is connecting this vertex has been
visited by particles of different classes, this particle might be absorbed
before reaching the vertex and then it will cease to affect the system.  If
the particle succeeds at moving to the neighboring vertex, it is said that
this particle has survived---and it will continue walking through the network.
This walking dynamics is modeled in terms of level of subordination and
domination of a class in relation to all other classes of particles.

In order to determine the level of domination and subordination of each
class in an edge we observe the active particles at a given time in the system.
Define \termdef{current directed domination} $\CurrentDomination{c}(t)$ to be
the number of active particles that belong to \class{c} that have decided to move
from \vertex{i} to \vertex{j} at \time{t} and survived.  Similarly,
define \termdef{current relative subordination} $\RelativeCurrentSubmission$ to
be the fraction of active particles that do not belong to \class{c} and have
successfully passed through \edge{i}{j} \emph{regardless of direction} at
time $t$.  The latter is defined as
\begin{displaymath}
    \RelativeCurrentSubmission \defeq \begin{dcases*}
        1 - \frac{
            \CurrentDomination{c} + \CurrentDomination[ji]{c}
        }{
            \sum_{q=1}^\PartClasses{\left(
                \CurrentDomination{q} + \CurrentDomination[ji]{q}
            \right)}
        } & if $\CurrentDomination{q} + \CurrentDomination[ji]{q} > 0, \forall q$\mbox{,} \\
        \frac{1}{\PartClasses} & otherwise.
    \end{dcases*}
\end{displaymath}

The survival of a particle depends on the current relative domination of the
edge and the destination vertex.  The survival probability is
\[
    1 - \SurvivalParam\RelativeCurrentSubmission(t)
\]
where $\SurvivalParam \in [0, 1]$ is the competition parameter.

Since particles are absorbed through the dynamics of competition, a mechanism
to perform replacement of absorbed particles is needed in order to avoid a
state where there are no active particles in the system.  This replacement
is done by creating new particles according to the distribution of the current
active particles of a given class in the system.

\newcommand{\InitialDiff}{ \left\lfloor\onpart[](0)-\onpart[](t)\right\rfloor }
Let $\npart[i](t)$ be the number of active particles that belong to \class{c}
at \time{t}.  The number of new particles that will belong to \class{c} in \vertex{i}
at \time{t} follows the distribution
\[
    \begin{dcases*}
        \mathrm{B}\!\left(\InitialDiff,~\Sourceness\right) & if $
        \left\lfloor
            \onpart[](0)-\onpart[](t)
        \right\rfloor > 0$, \\
        \mathrm{B}(1, 0) & otherwise,
    \end{dcases*}
\]
where
\[
    \Sourceness \defeq
        \frac{
          \onpart[i]
        }{
          \sum_{j=1}^{|V|}{
            \onpart[j]
            }
        }
\]
and $\mathrm{B}(n,p)$ is a binomial distribution.  In other words, if the
number of active particles is less than the initial number of active
particles, then it is performed $\InitialDiff$ trials with probability
$\Sourceness$ at generating a new particle in \vertex{i}.

It follows that the expected number of new particles that belong to
\class{c} in \vertex{i} at \time{t} is
\[
    \begin{dcases}
        \Sourceness\InitialDiff
        & \mbox{if } \InitialDiff > 0\mbox{,} \\
        0 & \mbox{otherwise.}
    \end{dcases}
\]

The information of the current directed domination $\CurrentDomination{c}(t)$
determines the class of particles that dominates each edge in the system.
The edges of the network are grouped in sets by the class that dominates them.
For each \class{c}, the subset of edges $\EdgeSet^c(t) \in \EdgeSet$ is
\[
    \EdgeSet^c(t) \defeq \left\{
        ij \middle\suchthat \argmax_q\!{
            \left(
            \onpartclass{q}(t) + \onpartclass[ji]{q} (t)
            \right) = c
        }
    \right\}\mbox{.}
\]

Define the unweighted subnetwork
\begin{equation}
    \label{eq:unfolding}
    G^c(t) \defeq \left(\VertexSet, \EdgeSet^c(t)\right)
\end{equation}
to be the \termdef{unfolding} of network $G$ according to \class{c} at
\time{t}.  This subnetwork can be interpreted as a subspace with
the most relevant relationships for a given class.   The available
information in these subnetworks will be utilized for the study of community detection and
data clustering.

Next, a formal modeling of this dynamical system is presented.

\subsection{Mathematical Modeling}

Formally, we define \ModelName as a dynamical \system{\tilde{X}(t)}.  Let
$\onpart[i](t)$ be the number of active particles that belong to \class{c} in
\vertex{i} at \time{t}.  The internal state of this dynamical system is
\begin{equation}
    \label{eq:osystem}
    \tilde{X}(t) \defeq \left[\,\ovpart(t)\,\right]^T\!\mbox{,}
\end{equation}
where
\[
    \ovpart(t) \defeq \big[\,\onpart[i](t)\,\big]_{i}\,\mbox{.}
\]

Let $\onewparticles(t)$ and $\tilde{a}^c_i(t)$ be,
respectively, the number of particles generated and absorbed in \vertex{i} at
\time{t}.  The evolution function~$\tilde\phi$ of the dynamical system is
\begin{displaymath}
  \tilde{\phi} : \begin{dcases}
        \onpart[i](t+1) =
        \onpart[i](t) + \sum_{j}{
            \left(\onpart[ji](t+1) - \onpart(t+1)\right)
        } +
        \onewparticles(t+1) - \tilde{a}_i^c(t+1) \, \mbox{.}
  \end{dcases}
\end{displaymath}

Intuitively, the number $\onpart[i]$ of active particles that are in a vertex is
the total number of particles arriving minus the number of particles leaving and
particles the have been absorbed. Additionally, there is a term for the number
$\onewparticles$ of generated particles.
 Values $\onpart$, $\onewparticles$, and $\tilde{a}^c_i$
are obtained \emph{stochastically} according to the dynamics of \WDynamicsName,
absorption, and generation.
The initial state of the system is given by an arbitrary number
$\onpart[i](0)$ of initial active particles and $\onpart[ij](0) = 0$.

In order to achieve the desirable network unfolding, it is necessary to average
the results of several simulations of the system with a very large number of
initial particles $\onpart[i](0)$.  However, the computational cost of a such
simulation is very high.  Alternatively, a \system{X(t)}
that achieves similar results in a \emph{deterministic} manner can be
modeled. This alternative
system considers that exists an asymptotically infinite number of initial active
particles.

\subsection{Alternative Mathematical Modeling}

Consider the dynamical system whose internal state is
\[
    X(t) \defeq \begin{bmatrix*}
        \vpart(t) = \big[\npart[i](t)\big]_i \\[0.1in]
        \Npart(t) = \big(\npart[ij](t)\big)_{ij} \\[0.1in]
    \end{bmatrix*}
\]
that is a nonlinear Markovian dynamical system with the deterministic
evolution function
\begin{equation}
    \label{eq:evolution}
    \phi\colon \begin{dcases}
        \vpart(t+1) = \vpart(t) \times \FinalTrans(X(t))
        \left( \vpart(t) \times \FinalTrans(X(t)) \cdot \vec{1} \right) ^ {-1} \\
        \Npart(t+1) = \diag{\vpart(t)} \times \FinalTrans(X(t)) \, \mbox{,} \\
    \end{dcases}
\end{equation}
where
\[
    \FinalTrans(X) \defeq \Big(\finaltrans(X)\Big)_{ij}\mbox{,}
\]
\begin{equation}
    \label{eq:prob}
        \finaltrans(X) \defeq
            \frac{w_{ij}}{ \sum_{k=1}^{|V|} w_{ik}}\left(
                1 - \SurvivalParam\submission(X)
                \right) \mbox{,}
\end{equation}
\begin{equation}
    \label{eq:subordination}
    \submission(X) \defeq \begin{dcases*}
    1 - \frac{
        \xnpart[ij]{c} + \xnpart[ji]{c}
    }{
        \sum_{q=1}^\PartClasses{\left(
            \xnpart[ij]{q} + \xnpart[ji]{q}
        \right)}
    } & if  $\xnpart[ij]{c} + \xnpart[ji]{c} > 0$, $\forall c$,\\
    \frac{1}{\PartClasses} & otherwise.
    \end{dcases*}%
\end{equation}

The initial state of \system{X} is given by an arbitrary discrete \emph{distribution}
$\vpart(0)$ of initial active particles and $\npart[ij](0) = 0$.
This system performs an unfolding in $\mathcal{O}(\PartClasses|\EdgeSet|)$ operations.

\subsection{Community Detection}

In the context of community detection, we simulate the alternative model with
$\PartClasses$ classes---here $\PartClasses$ is also the number of communities---up to \time{$T$} and
work with the unweighted subnetworks $G^c(T)$ in order
to determine the community structure within the network $G$.  A vertex is said to
belong to community~$c$ if the density of edges in a neighborhood
is higher in $G^c(T)$ than in every other unfolding. Formally, the community
to which \vertex{j} belongs to is
\[
  \Comm{j} = \argmin_{c\in\{1,\dots,\PartClasses\}}{
             S^{\text{density}}_{c,j}
    }\,\mbox{,}
\]
where $S^{\text{density}}_{c,j} \in [0,1]$ is
the grade of membership of $x_j$ on community~$c$.
Let $\mathcal{N}^o_{c,j}$ be the neighborhood of a given order $o$ of \vertex{j}
in the unfolding $G^c(T)$.  Denote the number of edges in this neighborhood
as $\card{\EdgeSet(\mathcal{N}^{o}_{c,j})}$.  Thus, the grade of membership
by density of \vertex{j} in unfolding $G^c(T)$ is
\begin{equation}
    \label{eq:sdensity}
    S^{\text{density}}_{c,j} \defeq\frac{
        \card{\EdgeSet(\mathcal{N}^{o}_{c,j})}
    }{
        \sum_{q=1}^{\PartClasses}{
            \card{\EdgeSet(\mathcal{N}^{o}_{q,j})}
        }
    }\,\mbox{.}
\end{equation}

\subsection{Data Clustering}

For the application of data clustering, the community structure itself can be
the dataset partition.  This might be enough for data that can be easily
partitioned. If the data, however, are nonlinearly distributed then a community
structure with the same number of communities as of classes might not yield
satisfactory results.  In such cases, disconnected groups of vertices within a
community can be the unfolding result.  Depending on the definition of
community in networks, treating a disconnected group as a single community
is an inconsistency.  For the problem of data clustering, I assume a cluster
of data point is not divided in two by a second cluster, and therefore this is the
case where disconnected groups within communities is an undesirable behavior.
  One way to circumvent this, is to superestimate the actual
number of clusters by imposing a little larger number $\PartClasses$ of classes of
particles.  After the end of the process, the community structure is reduced to
$\Classes$ communities, resulting thus in the final data clustering.

The reduction step tries to combine adjacent communities in such a way that
the network's modularity is maximized.  The modularity of a community structure
within a network is a quality function of how well-defined are the
communities~\cite{Newman2004}.  The intuition of modularity is that there should
be more links within a community than links among communities.  In this
aspect, the goal in the first step of the proposed technique is to simplify the
problem by creating an initial number of communities and then, in the second
step, the problem is treated as an optimization of the network's modularity.

Since there are ${\PartClasses \choose \Classes}$ possible combinations of
communities for the reduction, in this paper a greedy construction of the
dendrogram is performed.  Starting from the initial community structure given
by $\Comm{j}$, the merging of two adjacent communities that yields larger
modularity is sought out.
For a partition with $\PartClasses$ communities, the algorithm
tries all possible merges of adjacent communities that result in $\PartClasses -1$
communities.  As a consequence, up to ${\PartClasses \choose \PartClasses - 1}$
merges and modularity evaluations are performed.  Next, by picking the partition
with highest modularity, the next best merging is determined by performing up to
${\PartClasses - 1 \choose \PartClasses - 2}$ evaluations.  Repeat this until there are
$\Classes$ communities.  Thus, the maximum number of evaluated merges is
\[
  \sum_{i = \Classes}^{\PartClasses - 1} \! {i + 1 \choose i}
  = \frac{1}{2} \left( \PartClasses - \Classes \right) \left( \PartClasses + \Classes + 1 \right)
  = \Complexity{ \PartClasses ^ 2 - \Classes ^ 2 } = \Complexity{ \PartClasses ^ 2 }
  \mbox{.}
\]

Since only sparse networks do present a community structure~\cite{Fortunato2010}, the number of
adjacent communities and, hence, the number of merge evaluations are considerably lower.
Together with the process, the entire technique is simulated in the
order of $\mathcal{O}(\PartClasses|\EdgeSet| + \PartClasses^2)$.

\section{Computer Simulations} \label{sec:simulations}

In order to assess the proposed model, computer simulations were performed.
This section starts with simulations of the proposed model for the problem of
community detection in complex networks. Afterwards, the data clustering technique
is applied for both artificial and real-world datasets.

\subsection{Simulation for Community Detection}

A real-world network borrowed from social science literature is here utilized
for evaluating the ability to detect a community structure in a network.  The
chosen network is the Zachary's karate club~\cite{Zachary1977}, a very
well-known network that has become a benchmark test for community detection
methods.  This network describes the relationship of 34 members of a US
university club in the 1970s.  Each member is a node in the network, and a link
between two nodes indicates whether two members know each other.  As the time passed an internal
conflict between the president, John A., and the instructor, Mr.\ Hi, divided the
group between members that supported the president and the members that
supported the instructor.  The goal is to split the club members in a group
that supports the president and a group that supports the instructor.

\begin{figure}[!ht]
  \centering
  \includegraphics[width=.5\textwidth]{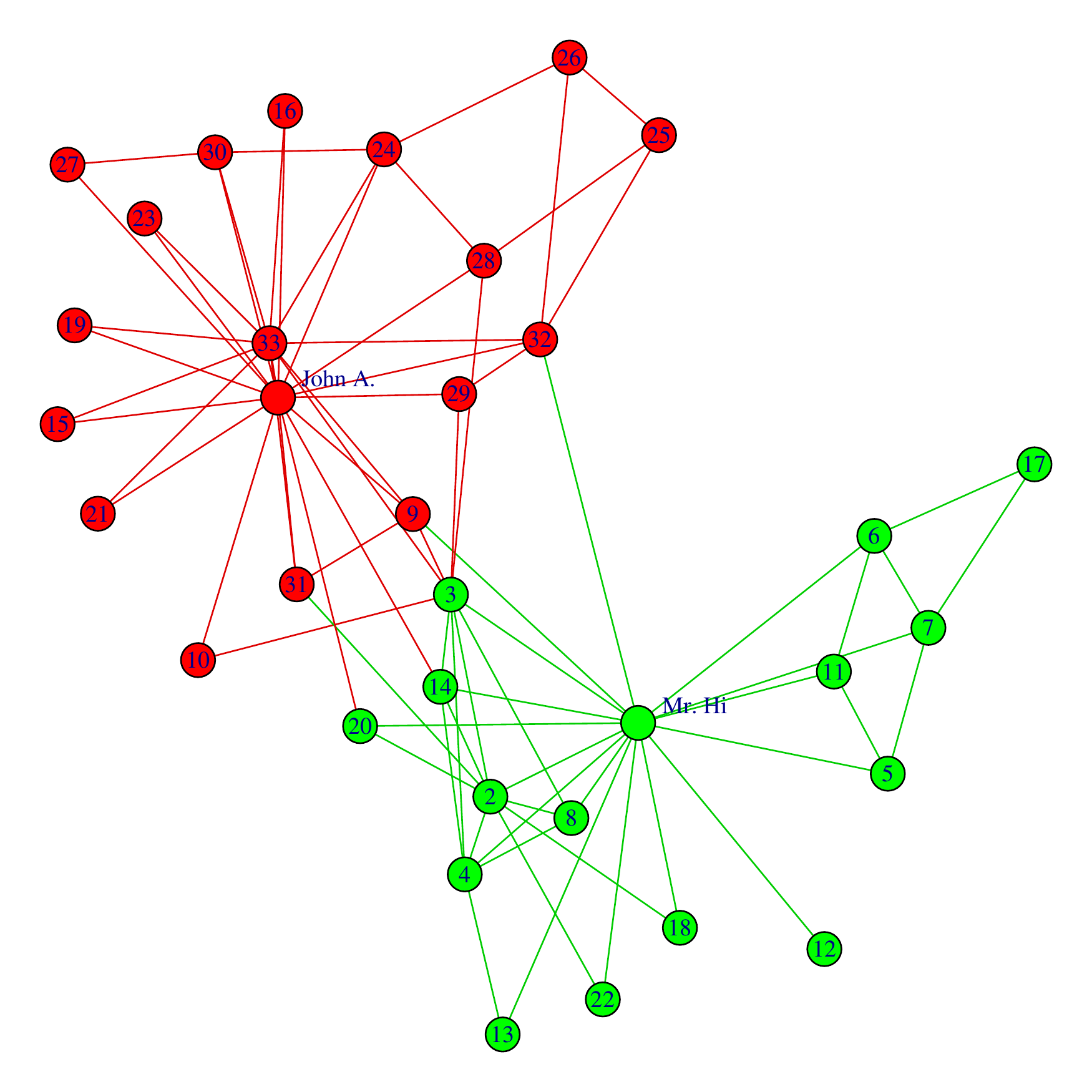}
  \caption{Community detection result for the proposed model on Zachary's
    karate club network.  Red and blue colors denote a group.  For edges
    the color is the class of particles that dominates the edge at the end of the
    simulation.  For vertices the color is the community to which a vertex
    has been grouped. All members were correctly grouped.}
    \label{fig:karate}
\end{figure}

In \cref{fig:karate}, the process' result for community detection on the
karate club is shown.  In edges, the colors represent the class of particles
that dominated the edge at the end of simulation.  In vertices, the color is
the result of community detection by using the density information.
The color of vertices represent the community structure detected by the
proposed model.  Even though this is an unweighted network (the links
denote only whether two members know each other) the proposed model was
able to detect the correct group of members by using the weight of 1 for
all links.

\subsection{Simulations for the Data Clustering Problem}

Six techniques with different approaches to tackle the problem of
data clustering were selected to compare with the proposed technique:
\kmeans~\cite{Hartigan1979},
Fuzzy \cmeans~\cite{Bezdek1984},
Hierarchical DBSCAN* with Density-Based Clustering Validation index (DBSCAN + DBCV)~\cite{Campello2013,Moulavi2014},
Chameleon~\cite{Karypis1999},
Expectation Maximization algorithm with parameterized Gaussian
mixture models~\cite{Fraley2012}, and
an adapted greedy algorithm based on modularity of networks~\cite{Clauset2004}.

In order to evaluate the quality of a partition found by an algorithm,
the measure of Adjusted Rand Index~\cite{Hubert1985} was calculated for
each generated partition. This measure index is defined in the range
$[-1, 1]$ where values close to 1 indicate that the partition is close to the
a prior knowledge of the dataset's clusters and values close to zero indicate
the partition is likely to be as good as an algorithm that partitions the
dataset at random.

Three techniques rely on stochastic initialization (\kmeans, fuzzy \cmeans,
and Expectation Maximization)---for these, the reported results are the average over 50 runs for each
dataset. For the simulations all techniques had prior knowledge
of the number of classes in the data clustering problem.  That way, the
reported results assess the ability of a technique to perform the dataset
partitioning given that the number of clusters is known.  In case of
Chameleon and the proposed technique, the algorithms start with an possibly
larger number of clusters/communities before agglomerating to the desirable
number.  For Chameleon, the initial number of clusters varied within
$\{\Classes, \Classes + 1, \dots, \card{\DataPoints}\}$, where
$\card{\DataPoints}$ is the number of data points in the dataset.
For the proposed technique, the competition parameter is fixed at $\SurvivalParam = 0.5$
with the number of classes of particles varying in
$\{2,  5, \dots, 30\}$. Moreover, modularity and the proposed technique are
graph-based methods, which implies the dataset input must be in a graph representation.
The \kNN method was used, varying $k \in \{1, 2, \dots, 30\}$.

\subsubsection{Simulations on Artificial Datasets}

\newcommand{\TblDatasetName}[1]{\bf #1}
\newcommand{\TblMethodName}[1]{\bf #1}
\newcommand{\TblResultValue}[2]{#1}
\newcommand{\TblResultRank}[1]{\bf #1}
\newcommand{\TblAvgRankName}[1]{\bf #1}

\begin{table}[t]
  \centering \scriptsize
  \caption{Adjusted Rand Index values for simulations on four artificial datasets.}
  \label{tbl:adjrand_artificial}
  \begin{tabular}{rrrrrr}
\toprule
&\TblDatasetName{Banana}&\TblDatasetName{Highleyman}&\TblDatasetName{Lithuanian}&\TblDatasetName{Spirals}&\TblAvgRankName{\pbox{20cm}{Avg.\\Rank}}\\
\midrule
\TblMethodName{\kmeans}&\TblResultValue{.2429}{.0172}&\TblResultValue{.2617}{.0099}&\TblResultValue{-.0016}{.0000}&\TblResultValue{-.0020}{.0000}&\TblResultRank{6.2}\\
\TblMethodName{Fuzzy \cmeans}&\TblResultValue{.2442}{.0016}&\TblResultValue{.3201}{.0000}&\TblResultValue{-.0017}{.0000}&\TblResultValue{-.0019}{.0000}&\TblResultRank{5.8}\\
\TblMethodName{HDBSCAN + DBCV}&\TblResultValue{.4714}{.0000}&\TblResultValue{.2085}{.0000}&\TblResultValue{.7024}{.0000}&\TblResultValue{.2507}{.0000}&\TblResultRank{4}\\
\TblMethodName{Chameleon}&\TblResultValue{.9215}{.0000}&\TblResultValue{.4348}{.0000}&\TblResultValue{.9343}{.0000}&\TblResultValue{.0119}{.0000}&\TblResultRank{3}\\
\TblMethodName{Expectation Maximization}&\TblResultValue{.3304}{.0505}&\TblResultValue{.7977}{.0000}&\TblResultValue{-.0015}{.0000}&\TblResultValue{-.0020}{.0000}&\TblResultRank{4.5}\\
\TblMethodName{Modularity}&\TblResultValue{.3510}{.0000}&\TblResultValue{.5033}{.0000}&\TblResultValue{.4259}{.0000}&\TblResultValue{1.0000}{.0000}&\TblResultRank{3}\\
\TblMethodName{\emph{Proposed technique}}&\TblResultValue{.9408}{.0000}&\TblResultValue{.7164}{.0000}&\TblResultValue{.9538}{.0000}&\TblResultValue{1.0000}{.0000}&\TblResultRank{1.5}\\
\bottomrule
\end{tabular}

\end{table}

Four artificial datasets were generated with PRTools framework~\cite{Duin2007}.
The datasets are formed by points in two-dimensional space equally splitted
into two classes that form two clusters to be detected by the algorithm. Except
for the dataset of spirals that contain 500 points, the other four datasets are
formed by 600 points.  Because the datasets are not in a network
representation, the \emph{k}-Nearest Neighbor (\kNN) graph construction method
is employed to obtain a graph that is the system's network. In the constructed
network, a vertex represents a data point, and it connects to its $k$ nearest
points determined by Euclidean distance.

\cref{fig:banana,fig:lithuanian,fig:highleyman,fig:spirals} show the
generated dataset, the result obtained by application of \kmeans, and the
result of the proposed technique.  In \cref{tbl:adjrand_artificial}, the
adjusted rand index of each technique is shown.  The last column is the
average ranking position that a technique obtained.  The technique is
able to correctly detect the shape distribution in the four datasets.
The graph representation naturally determine how particles flow through the
network. If the clusters have a slow interconnectivity, as the case of
spirals (see \cref{fig:spirals}), both modularity and the proposed,
which both are network-based techniques, are able to correctly partition the
data without any mistake.  But the network is not the only important aspect.
For instance, modularity could not partition data points of different classes
in overlapping regions, as the case of Highleyman dataset, whereas the proposed
was able to find a disconnected community (see \cref{fig:highleyman}).

\newcommand{\Scale}{0.5\textwidth}

\begin{figure}
  \begin{minipage}[t]{.45\textwidth}
  \subfigure[]{%
    \includegraphics[width=\Scale]{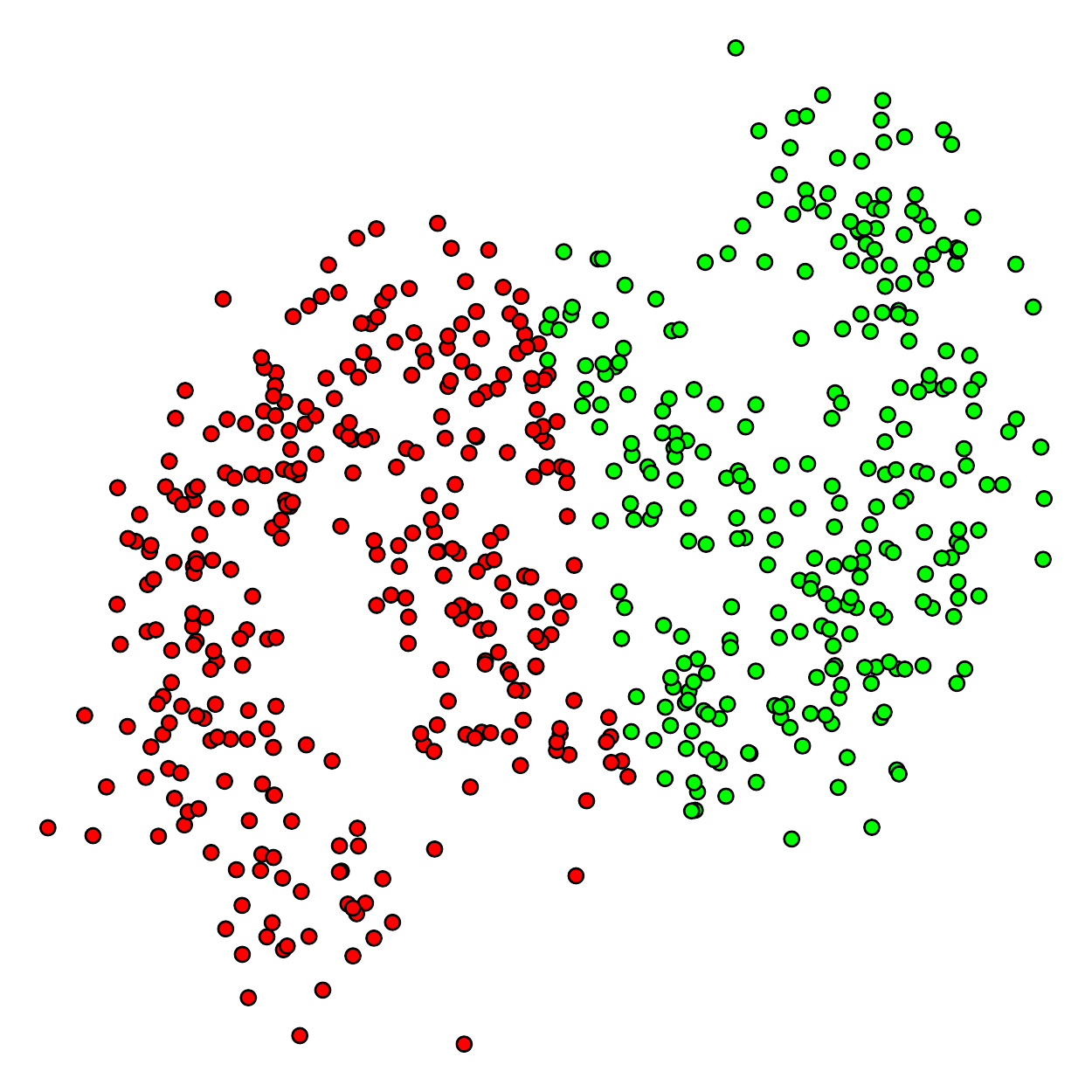}%
  }%
  \subfigure[]{%
    \includegraphics[width=\Scale]{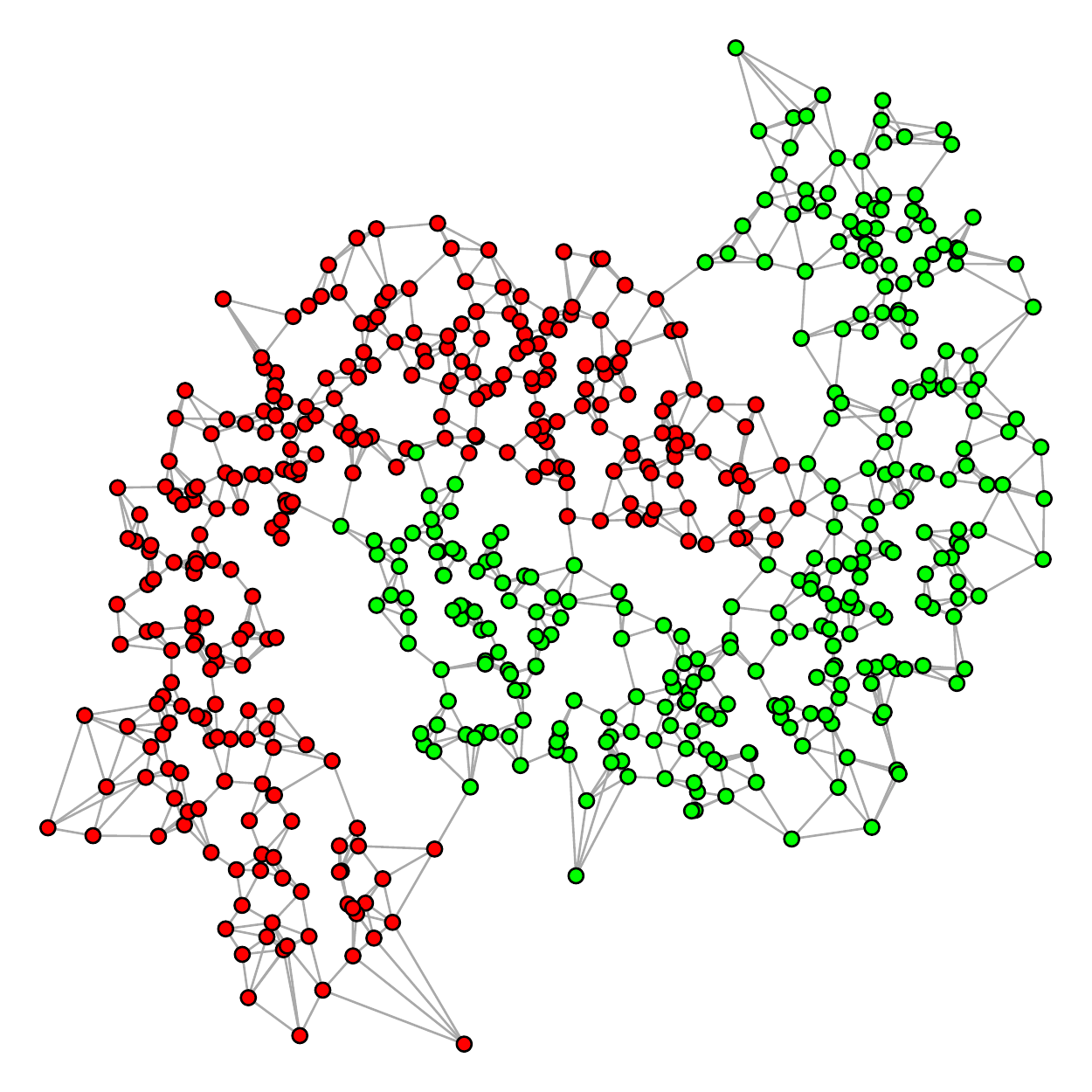}
  }%
  \caption{Banana-shaped distribution dataset with two classes.
    (a) Result obtained by \kmeans.
    (b) Result obtained by the proposed technique.
  }
  \label{fig:banana}
  \end{minipage}\hfill%
  \begin{minipage}[t]{.5\textwidth}
  \subfigure[]{%
    \includegraphics[width=\Scale]{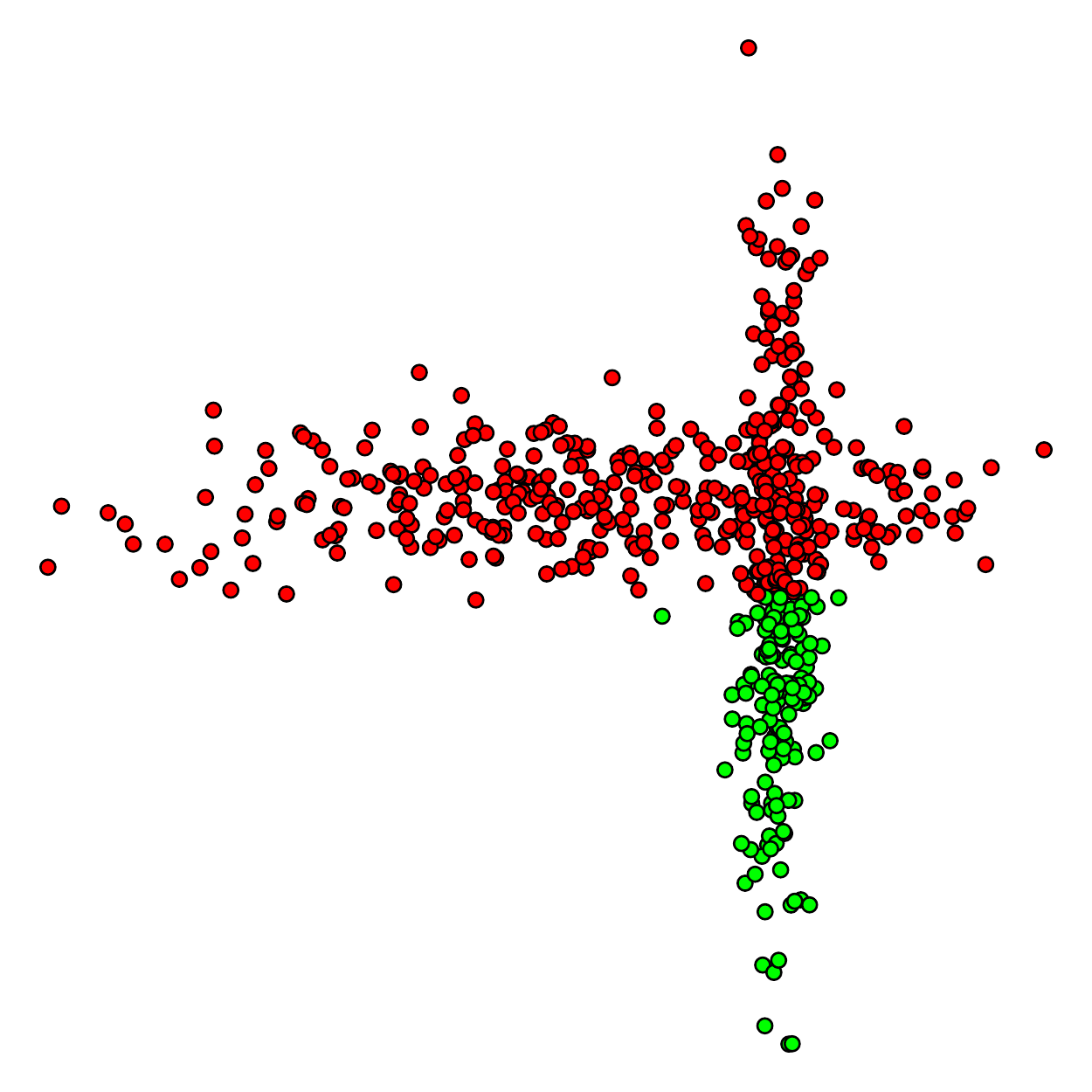}%
  }%
  \subfigure[]{%
    \includegraphics[width=\Scale]{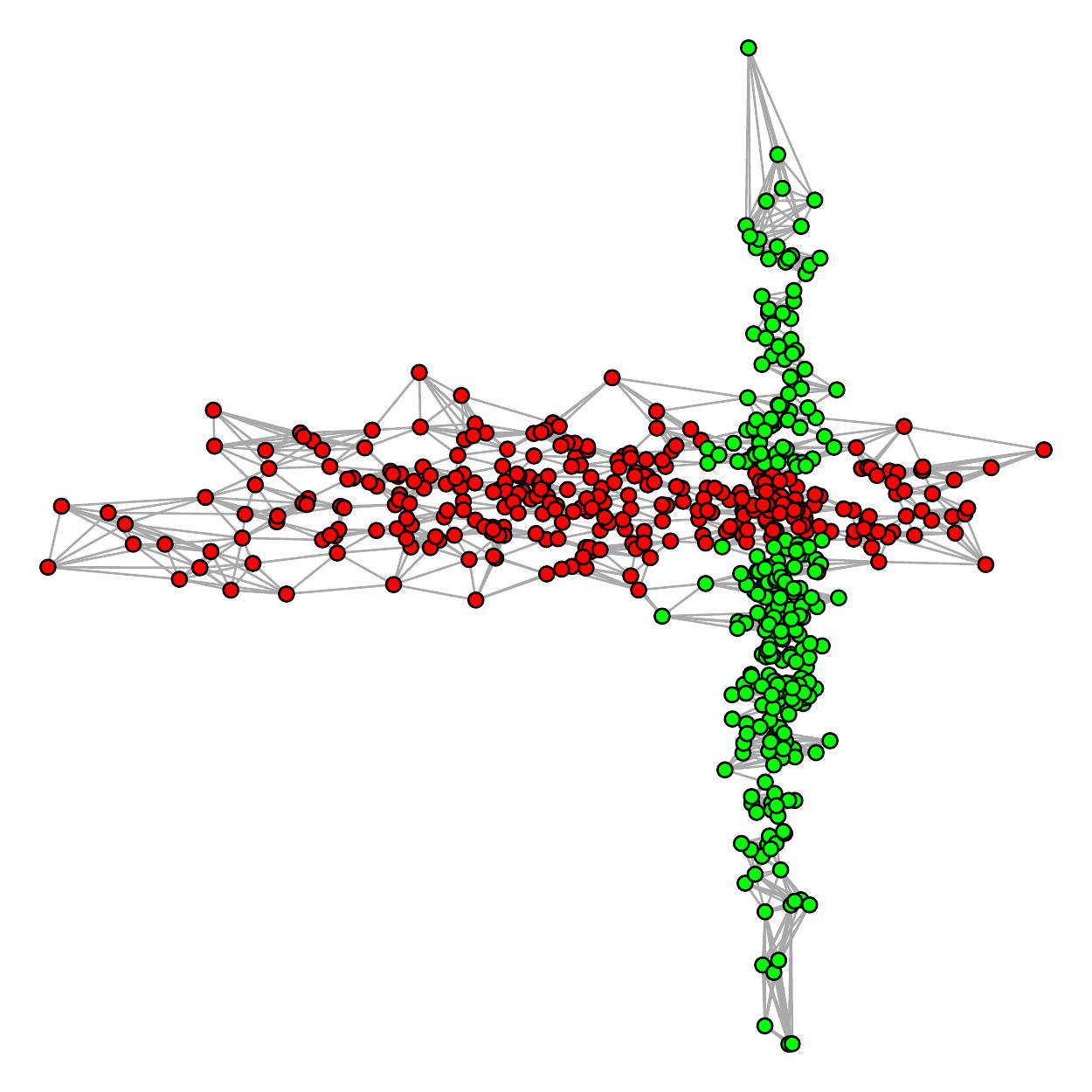}
  }%
  \caption{Highleyman dataset with two classes.
    (a) Result obtained by \kmeans.
    (b) Result obtained by the proposed technique.
  }
  \label{fig:highleyman}
\end{minipage}
\end{figure}

\begin{figure}
  \begin{minipage}[t]{.47\textwidth}
  \subfigure[]{%
    \includegraphics[width=\Scale]{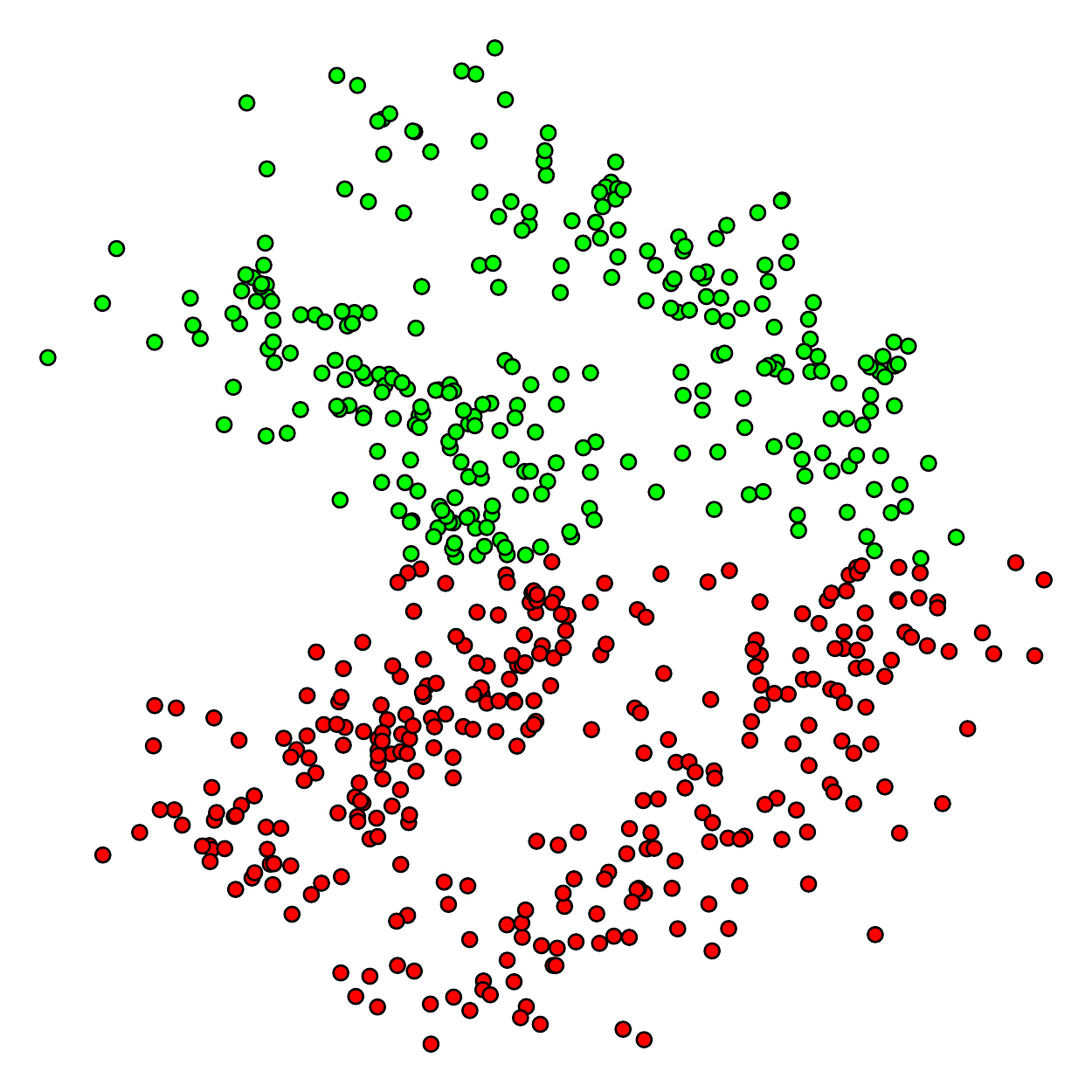}%
  }%
  \subfigure[]{%
    \includegraphics[width=\Scale]{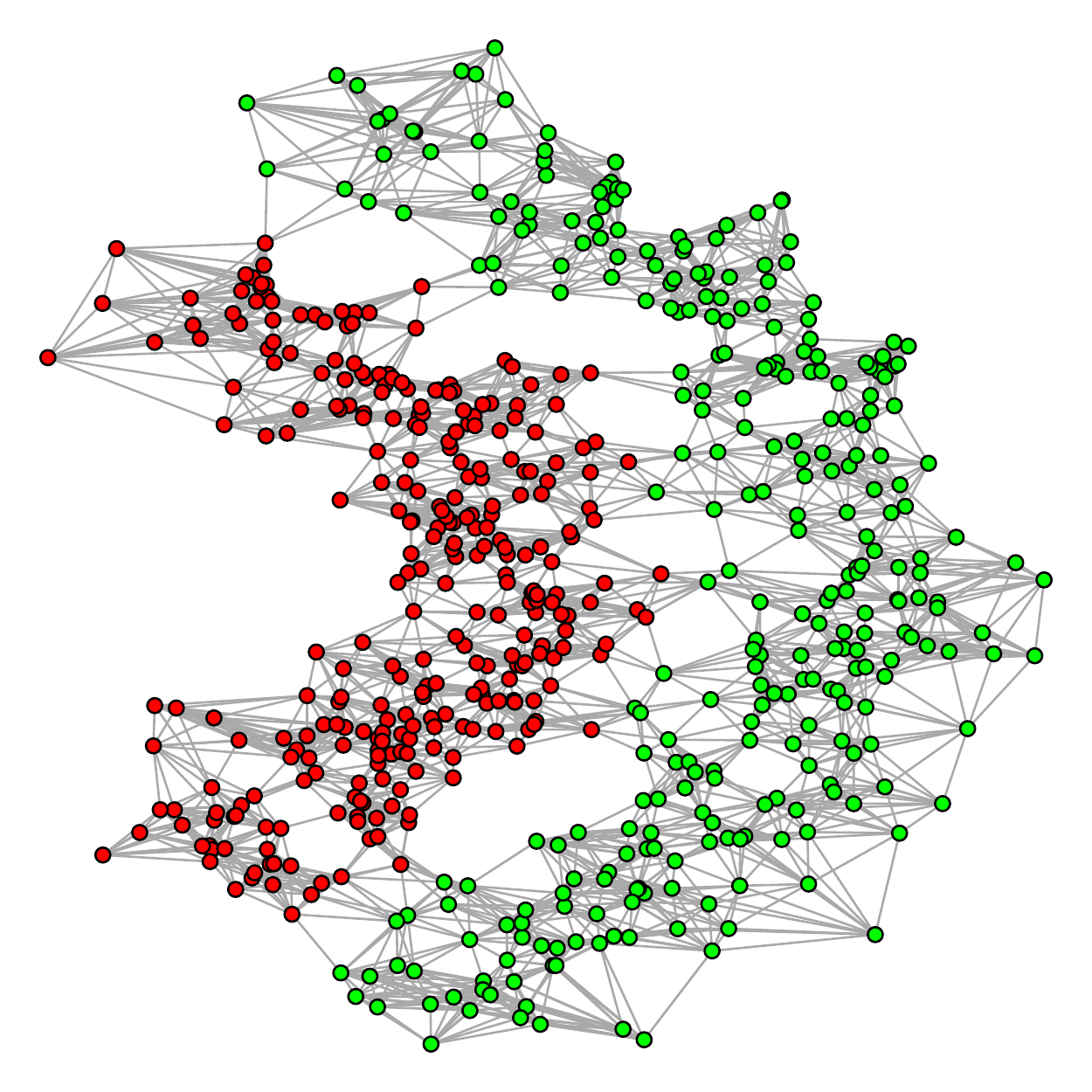}
  }%
  \caption{Lithuanian dataset with two classes.
    (a) Result obtained by \kmeans.
    (b) Result obtained by the proposed technique.
  }
  \label{fig:lithuanian}
  \end{minipage}\hfill%
  \begin{minipage}[t]{.5\textwidth}
  \subfigure[]{%
    \includegraphics[width=\Scale]{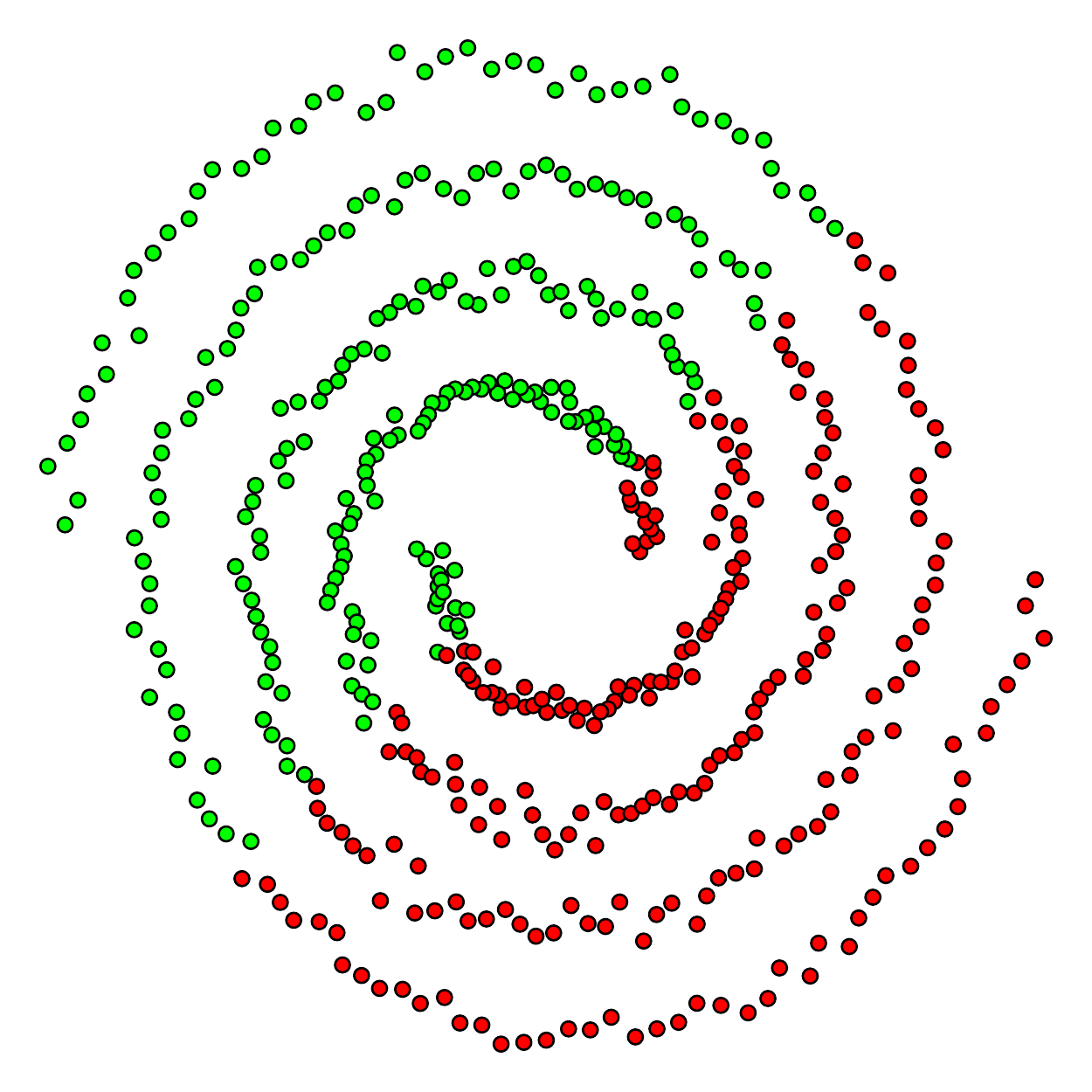}%
  }%
  \subfigure[]{%
    \includegraphics[width=\Scale]{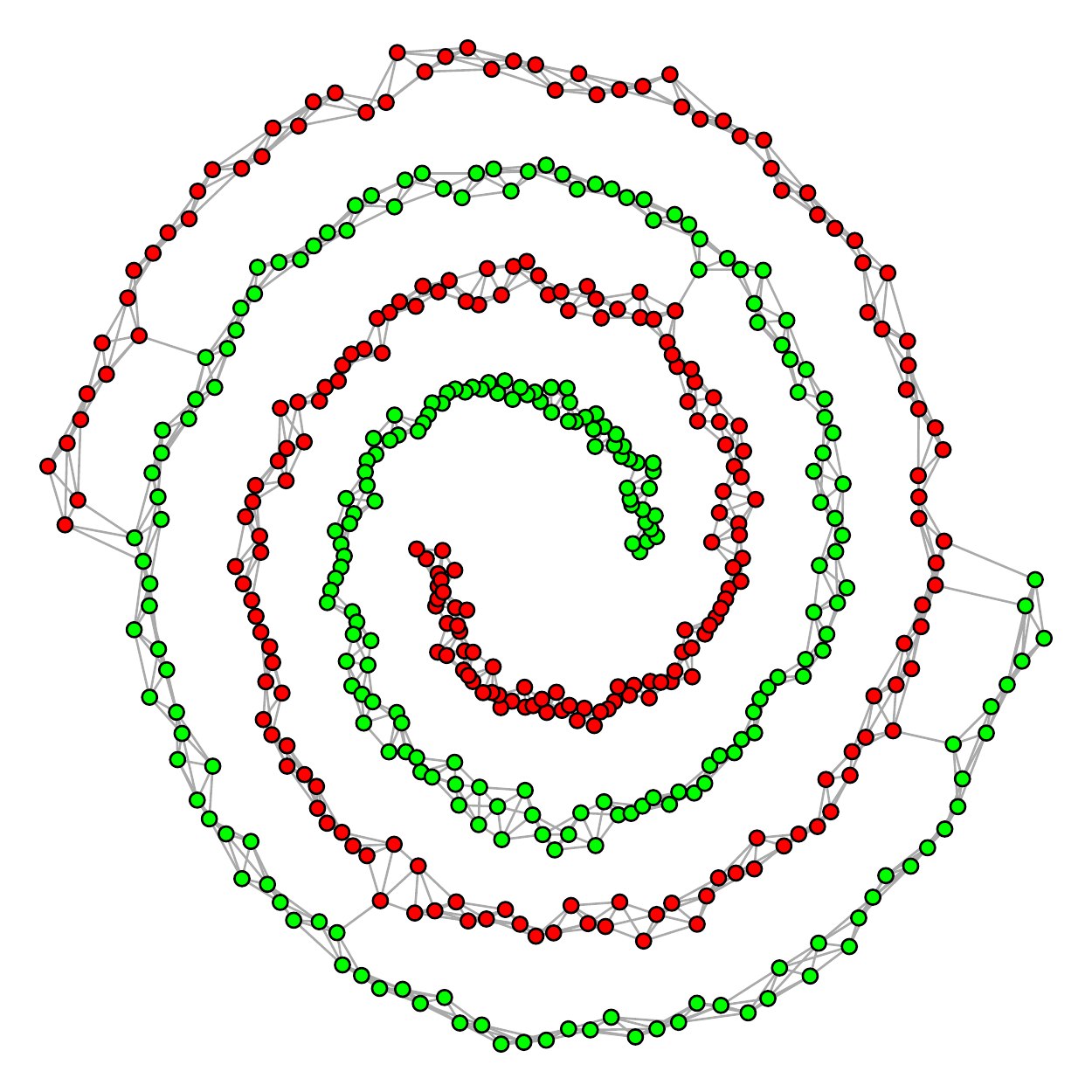}
  }%
  \caption{Spirals dataset with two classes.
    (a) Result obtained by \kmeans.
    (b) Result obtained by the proposed technique.
  }
  \label{fig:spirals}
\end{minipage}
\end{figure}


\subsubsection{Simulations on Real Datasets}

Ten real-world datasets were selected from the UCI repository~\cite{Asuncion2007}.
A brief description of the datasets is given in \cref{tbl:descrealdatasets}.
Results of adjusted rand index for simulations of these datasets is shown
in~\cref{tbl:adjrand_real}.  In this table, the last column is the
average ranking position of each technique over all datasets.  The ranking
is sorted accordingly to the adjusted rand index. A higher index indicates
that a technique has partitioned the dataset in a way more similar to the
prior knowledge.  Thus, the technique that obtained the highest index for
a dataset has a ranking position of 1 for that dataset; the second highest index
is associated to a ranking position of 2, and so on.

The parameter combination the obtained the highest adjusted rand index
is shown in~\cref{tbl:adjrand_parameters}.
Notably for all real-world datasets, simulating the system with a higher value
for the number $\PartClasses$ of classes of particles, and thereafter in a
second step merging to $\Classes$ communities has given the best results.

\begin{table}[ht]
  \centering \scriptsize
  \caption{Description of the real-world datasets, with the number of
  data points, the number of dimensions in the attribute space, and the
  number of clusters the data points are partitioned.}
  \label{tbl:descrealdatasets}
  \begin{tabular}{lrrr}
    \toprule
    & \bf Instances & \bf Dimensions & \bf Classes \\
    \midrule
\bf Breast cancer        & 699   &  9   &  2  \\
\bf Car evaluation       & 1728  &  6   &  4  \\
\bf Credit approval      & 690   &  15  &  2  \\
\bf Contraceptive method & 1473  &  9   &  3  \\
\bf Glass                & 214   &  9   &  6  \\
\bf Ionosphere           & 351   &  34  &  2  \\
\bf Iris                 & 150   &  4   &  3  \\
\bf Vowel                & 90    &  10  &  11 \\
\bf Wine                 & 178   &  13  &  3  \\
\bf Seeds                & 270   &  7   &  3  \\
    \bottomrule
  \end{tabular}
\end{table}

\begin{table}[t]
  \centering \scriptsize
  \caption{Comparison of the Adjusted Rand Index of seven different techniques
  on 10 real-world datasets.}
  \label{tbl:adjrand_real}
  \begin{adjustbox}{width={\textwidth}}
    \tiny
    \begin{tabular}{rrrrrrrrrrrr}
\toprule
&\TblDatasetName{\pbox{20cm}{Breast\\cancer}}&\TblDatasetName{\pbox{20cm}{Car\\eval.}}&\TblDatasetName{\pbox{20cm}{Credit\\approval}}&\TblDatasetName{\pbox{20cm}{Contr.\\method}}&\TblDatasetName{Glass}&\TblDatasetName{Ion.}&\TblDatasetName{Iris}&\TblDatasetName{Vowel}&\TblDatasetName{Wine}&\TblDatasetName{Seeds}&\TblAvgRankName{\pbox{20cm}{Avg.\\Rank}}\\
\midrule
\TblMethodName{\kmeans}&\TblResultValue{.7302}{.0000}&\TblResultValue{.0294}{.0276}&\TblResultValue{.2389}{.1755}&\TblResultValue{.0215}{.0058}&\TblResultValue{.1610}{.0223}&\TblResultValue{.1776}{.0000}&\TblResultValue{.6540}{.1186}&\TblResultValue{.1736}{.0158}&\TblResultValue{.8582}{.0135}&\TblResultValue{.7049}{.0000}&\TblResultRank{4.8}\\
\TblMethodName{Fuzzy \cmeans}&\TblResultValue{.7305}{.0000}&\TblResultValue{.0307}{.0265}&\TblResultValue{.3725}{.0005}&\TblResultValue{.0242}{.0000}&\TblResultValue{.1632}{.0054}&\TblResultValue{.1727}{.0000}&\TblResultValue{.7287}{.0000}&\TblResultValue{.0892}{.0209}&\TblResultValue{.8498}{.0000}&\TblResultValue{.7266}{.0000}&\TblResultRank{4.3}\\
\TblMethodName{HDBSCAN+DBCV}&\TblResultValue{.2556}{.0000}&\TblResultValue{.1313}{.0000}&\TblResultValue{.0794}{.0000}&\TblResultValue{.0236}{.0000}&\TblResultValue{.2575}{.0000}&\TblResultValue{.7030}{.0000}&\TblResultValue{.5657}{.0000}&\TblResultValue{.0814}{.0000}&\TblResultValue{.3385}{.0000}&\TblResultValue{.4303}{.0000}&\TblResultRank{5.4}\\
\TblMethodName{Chameleon}&\TblResultValue{.7192}{.0000}&\TblResultValue{.1496}{.0000}&\TblResultValue{.1653}{.0000}&\TblResultValue{.0253}{.0000}&\TblResultValue{.2918}{.0000}&\TblResultValue{.6767}{.0000}&\TblResultValue{.6844}{.0000}&\TblResultValue{.1949}{.0000}&\TblResultValue{.8249}{.0000}&\TblResultValue{.7436}{.0000}&\TblResultRank{3.6}\\
\TblMethodName{E.~Maximization}&\TblResultValue{.6955}{.0941}&\TblResultValue{.0367}{.0454}&\TblResultValue{.1987}{.1741}&\TblResultValue{.0112}{.0097}&\TblResultValue{.1571}{.0234}&\TblResultValue{.1547}{.0273}&\TblResultValue{.9222}{.0000}&\TblResultValue{.1541}{.0172}&\TblResultValue{.9472}{.0000}&\TblResultValue{.6671}{.0454}&\TblResultRank{4.8}\\
\TblMethodName{Modularity}&\TblResultValue{.4474}{.0000}&\TblResultValue{.1872}{.0000}&\TblResultValue{.1734}{.0000}&\TblResultValue{.0329}{.0000}&\TblResultValue{.2118}{.0000}&\TblResultValue{.0708}{.0000}&\TblResultValue{.9038}{.0000}&\TblResultValue{.2505}{.0000}&\TblResultValue{.8858}{.0000}&\TblResultValue{.8125}{.0000}&\TblResultRank{3.5}\\
\TblMethodName{\emph{Proposed technique}}&\TblResultValue{.7930}{.0000}&\TblResultValue{.1880}{.0000}&\TblResultValue{.4890}{.0000}&\TblResultValue{.0433}{.0000}&\TblResultValue{.2377}{.0000}&\TblResultValue{.3057}{.0000}&\TblResultValue{.9222}{.0000}&\TblResultValue{.2259}{.0000}&\TblResultValue{.9488}{.0000}&\TblResultValue{.8377}{.0000}&\TblResultRank{1.6}\\
\bottomrule
\end{tabular}

  \end{adjustbox}
\end{table}

\begin{table}[t]
  \centering
  \caption{Best parameter combinations for both artificial and real-world
    datasets. The best combination is associated to the
    highest adjusted rand index obtained. The listed parameters are the
  \kNN, the number of classes of particles, and the order of the neighborhood
  considered for grouping vertices in communities.}
    \label{tbl:adjrand_parameters}
    \scriptsize
  \begin{tabular}{lrrr}
    \toprule
     & \bf \kNN & \bf $\PartClasses$ & \bf Neighborhood \\
    \midrule
\bf Banana shape           & 4& 2&4\\
\bf Lithuanian             & 7&24&2\\
\bf Highleyman             &11& 8&4\\
\bf Spirals                & 5&18&1\\ \hline
\bf Breast cancer          & 4&30&3\\
\bf Car evaluation         &16&11&4\\
\bf Credit approval        &28&27&3\\
\bf Contraceptive method   &10&11&2\\
\bf Glass                  &28& 8&3\\
\bf Ionosphere             & 9&30&2\\
\bf Iris                   & 8& 5&1\\
\bf Vowel                  & 7&30&2\\
\bf Wine                   & 8&24&3\\
\bf Seeds                  & 4&21&3\\
    \bottomrule
  \end{tabular}
\end{table}

It is expected that the modularity algorithm do not have a significant
difference, since the intuition of the present technique is to simplify
the problem by finding some communities in a first step for later optimize
the partitions according to a quality function, such as the modularity.
An interesting aspect of the reported results is that the average rank
of presented technique do not vary much accordingly to the dataset domain,
which could mean the technique does not have a great bias of domain.
Nevertheless, the artificial toy datasets have a show case where the
technique is able to get a better partitioning by loosing the definition
of community and allowing disconnected communities.

Chameleon has a very similar insight as the proposed technique in the way it
approaches the data clustering problem.  First, Chameleon splits the network
representation of the dataset into small groups of vertices. Afterwards,
it agglomerates these groups into larger ones until obtained the
desired number of clusters.  One problem is the step of defining the
small groups. If the small groups of vertices are not well defined---such
as in cases of data with overlapping regions---, the agglomerative step will
result in a poor data clustering.  Conversely, if well-defined then it
obtains good results as in the case of the Glass and Ionosphere datasets,
which were the two datasets the Chameleon obtained a higher value for
adjusted rand index.

\begin{figure}[t]
  \centering
  \includegraphics[width=\textwidth,keepaspectratio]{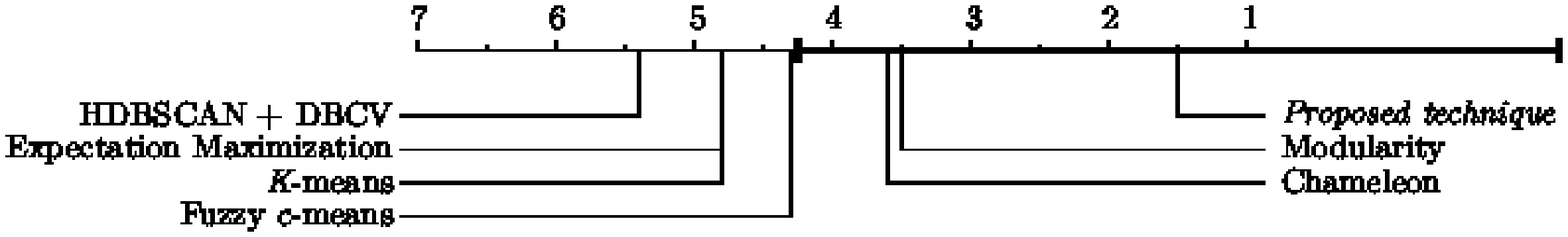}
  \caption{Visualization of post-hoc Bonferroni--Dunn test.  Fixing
    the proposed technique as the control, the average ranking of the other
    six techniques were compared.
    All techniques with average rankings
  outside the bold-marked interval are significantly different from the
  proposed technique, with a significance level of $5\%$.}
  \label{fig:posthoc}
\end{figure}

In order to evaluate whether the average ranking position of the techniques
are significantly different, the average rankings are examined in a statistical
manner.  The methodology applied here follows a procedure described
in~\textcite{Demsar2006,Japkowicz2011}. The first test, Friedman test compares the techniques
under the null-hypothesis that all the techniques are equivalent and so their
average rank positions.  According to \textcite{Demsar2006}, since we have
$N = 10$ datasets and $k = 7$ techniques, we have the degrees of
freedom $\text{\itshape df1} = k - 1 = 6$ and
$\text{\itshape df2} = (N - 1) (k - 1) = 54$, resulting in
the critical value $F(6, 54) \approx 2.27$.  With such values, we get a
Friedman statistic $F_F \approx 4.88$.  Since $F_F > F(6, 54)$ the
null-hypothesis is rejected at a $5\%$ significance level, allowing us to
proceed with a post-hoc test.

Fix a significance level of $5\%$ for the post-hoc.  The interest is at
comparing the proposed model with the other six techniques.  For that the
Bonferroni--Dunn test is employed, because there are multiple techniques
over multiple datasets, and only the difference between a fixed technique and
all others is of interest~\cite{Japkowicz2011}.  Therefor, six
null-hypothesis of the proposed technique's average rank being equivalent
to an other technique are tested.  To accomplish this, the $t$ statistic
between a fixed and an other technique must be by at least a Critical
Difference (CD).  If they differ that much, then the null-hypothesis is
rejected and therefore there is a significant statistical difference between
the average ranking position of the two techniques. For the setting in this
paper, we get a $\text{CD} \approx 3.33$.
The proposed technique has significant statistical difference when
compare with \kmeans, Fuzzy \cmeans, HDBSCAN + DBCV, and Expectation
Maximization.  The techniques Chameleon and Modularity do not present
significant statistical difference. But, nevertheless, these two together
with the proposed technique, put network-based methods in a good position
at data clustering problems.
A visualization of the post-hoc test, as suggested
in~\cite{Demsar2006}, is shown in~\cref{fig:posthoc}.

\section{Conclusion} \label{sec:conclusions}

In this paper, a novel model for particle competition in complex networks
is presented.  Particles flow through the network and their presence dominates
an edge.  The information is then used to determine the pertinence of vertices
in communities of the network.  A second step allows the maximization of
the modularity by locally maximizing the modularity of the community structure.
This final structure is used to the problem of data clustering.

Simulations on both artificial and real-world datasets show that the
proposed technique performs well and have a significant better performance
under the condition that the algorithms have prior knowledge of the number
of clusters.
Furthermore, the technique has a deterministic alternative
model with a low computational complexity.

The information of dominance of edges have a potentially higher granularity
information about the interaction network formed by the graph representation of
the dataset. This information might allow to grasp more properties of the
process and consequently help into find a better approach to the local
maximization of the community structure.

Finally, in a data clustering setting it is not uncommon to run algorithms on
datasets with noise and no prior knowledge of the number of clusters.  The
proposed technique here present could be extended for such applications.

\printbibliography

\end{document}